\documentstyle[12pt]{article}
\newcommand{\thspace}{\kern.08333em}

\def \beq{\begin{equation}}
\def \eeq{\end{equation}}
\def \s{\sqrt{2}}
\def\ubar{\overline{u}}
\def\cbar{\overline{c}}

\def\dbar{\overline{d}}
\def\sbar{\overline{s}}
\def\bbar{\overline{b}}
\def\fbar{\overline{f}}
\def\qbar{\overline{q}}

\def\Abar{\overline{A}}
\def\Kbar{\overline{K}^0}
\def\Bbar{\overline{B}^0}

\def\bd{B^0}
\def\bo{B^0}
\def\bs{B_s}
\def\bdb{\overline{B}^0}
\def\bsb{\overline{B}_s}

\def\bq{B_q}
\def\bqb{\bar{B_q}}
\def\Dbar{\overline{D}^0}

\def\Gammabar{\overline{\Gamma}}
\def\to{\rightarrow}
\begin{document}
%
\rightline{TECHNION-PH-97-17}
\rightline{May 1997}
\bigskip
\bigskip
\centerline {\bf CP VIOLATION IN THE $B$ MESON SYSTEM~\footnote
{Invited talk given at FCNC97, Santa Monica, CA, Feb. 19$-$21, 1997}}
\vskip 2.0cm
\centerline {MICHAEL GRONAU}
\centerline {\it Department of Physics}
\centerline {\it Technion - Israel Institute of Technology, 32000 Haifa, Israel}
\vskip 3.0cm
\centerline {\bf Abstract}
\vskip 1.0cm
\noindent
CP violation in $B$ decays is reviewed in the Standard Model (SM)
and beyond the SM. The present explanation of CP violation in terms
of a phase in the Cabibbo-Kobayashi-Maskawa (CKM) matrix can be tested through
a variety of CP asymmetries in neutral and charged $B$
decays. Usually, new mechanisms of CP nonconservation enter via $B-\overline{B}$
mixing and violate SM constraints on the CKM parameters in a few characteristic
ways.  Different models can be  partially distinguished by penguin-dominated
$B$ decay rate measurements. In radiative decays, large mixing-induced
asymmetries may occur due to new contributions to the decay amplitude.

\vfill

\newpage
\section{Introduction: The CKM Matrix}
In the Standard Model (SM), CP violation is
due to a nonzero complex phase in the Cabibbo-Kobayashi-Maskawa (CKM) matrix
$V$, describing the interaction of the three families of
quarks with the charged gauge boson. This unitary matrix can be approximated
by the following two useful forms \cite{PDG}:
$$
V \approx \left(\matrix{1-{1\over 2}s_{12}^2&s_{12}&s_{13}e^{-i\gamma}\cr
-s_{12}&1-{1\over 2}s_{12}^2&s_{23}\cr
s_{12}s_{23}-s_{13}e^{i\gamma}&-s_{23}&1\cr}\right)
$$
\beq \label{V}
\approx \left(\matrix{1-{1\over 2}\lambda^2&\lambda&A\lambda^3(\rho-i\eta)\cr
 -\lambda&1-{1\over 2}\lambda^2&A\lambda^2\cr
A\lambda^3(1-\rho-i\eta)&-A\lambda^2&1\cr}\right)~.
\eeq
The measured values of the three inter-generation mixing angles , $\theta_{ij}$,
and the phase $\gamma$ are given by \cite{ALON}:
$$
s_{12}\equiv \sin\theta_{12}\approx\vert V_{us}\vert=0.220\pm 0.002~,
$$
$$
s_{23}\equiv \sin\theta_{23}\approx\vert V_{cb}\vert=0.039\pm 0.003~,
$$
$$
~~~~s_{13}\equiv \sin\theta_{13}\equiv\vert V_{ub}\vert=0.0031\pm 0.0008~,
$$
\beq \label{ANGLES}
35^0\leq\gamma\equiv{\rm Arg}(V^*_{ub})\leq 145^0~.
\eeq
The only information about a nonzero value of $\gamma$
comes from CP violation in the $K^0-\Kbar$ system.

\noindent
Unitarity of $V$ implies quite a few triangle relations. The $db$ triangle,
\beq\label{UNIT}
V_{ud}V^*_{ub}+V_{cd}V^*_{cb}+V_{td}V^*_{tb}=0~,
\eeq
which has large angles, is shown in the latest Review of Particle Physics
\cite{PDG}.
The phase $\beta={\rm Arg}(V^*_{td})$ is determined to lie within the limits
\beq\label{BETA}
10^0\leq\beta\leq 35^0~,
\eeq
while the present bounds on the third angle, $\alpha\equiv \pi/2-\beta-\gamma$,
are
\beq\label{ALPHA}
20^0\leq\alpha\leq 120^0~.
\eeq
In addition to the separate constraints on $\alpha,~\beta$ and $\gamma$, pairs
of these angles are correlated. Due to the rather limited range of $\beta$, the
angles $\alpha$ and $\gamma$ are almost linearly correlated through ~$\alpha +
\gamma = \pi- \beta$ \cite{DGR}.
A special correlation exists also between small values of $\sin 2\beta$ and
large values of $\sin 2\alpha$ \cite{NISAR}.

\noindent
A precise determination of the three angles $\alpha,~\beta$ and $\gamma$,
which would provide a test of the CKM origin of CP violation, relies on
measuring CP asymmetries in $B$ decays. A few methods, which by now became
standard, are described shortly in Section 2. This discussion includes
a new variant of a method which determines $\gamma$ from charged $B$ decays. The
use of flavor SU(3) and first-order SU(3) breaking in analyzing $B$ decays to
two pseudoscalar mesons is the subject of Section 3. We will point out the
importance of the large branching ratio measured recently for $B^+\to\eta'
K^+$.
Section 4 reviews CP violation in the $B$ meson system beyond the SM. We will
demonstrate the complementary role played by CP asymmetries, on the one hand,
and rare penguin $B$-decays, on the other hand, in distinguishing among
different models of CP violation. An interesting and quite unusual
mechanism will be shown to lead in some models to large CP asymmetries
in radiative neutral $B$ decays. A brief conclusion with a future outlook is
given in Section 6. More details on some of these methods and further
references can be found in previous reviews \cite{REV}.

\section{Methods of Measuring CKM Phases}

\subsection{\it Decays to CP-eigenstates}

The most frequently discussed method of measuring weak phases is based on
neutral
$B$ decays to final states $f$ which are common to $B^0$ and $\Bbar$. CP
violation is induced by $B^0-\Bbar$ mixing through the interference of the two
amplitudes $B^0\to f$ and $B^0\to\Bbar\to f$. When $f$ is a CP-eigenstate, and
when a single weak amplitude (or rather a single weak phase) dominates the
decay process, the time-dependent asymmetry
\beq
{\cal A}(t)\equiv {\Gamma(B^0(t)\to f)-\Gamma(\Bbar (t)\to f)\over
\Gamma(B^0(t)\to f)+\Gamma(\Bbar (t)\to f)}
\eeq
obtains the simple form \cite{BIGSAN}
\beq\label{ASYM}
{\cal A}(t)= \xi\sin2(\phi_M+\phi_f)\sin(\Delta mt)~.
\eeq
$\xi$ is the CP eigenvalue of $f$, $2\phi_M$ is the phase of $B^0-\Bbar$ mixing,
($\phi_M=\beta,~0$ for $B^0_d,~B^0_s$, respectively), $\phi_f$ is the weak
phase of the $B^0\to f$ amplitude, and $\Delta m$ is the neutral $B$
mass-difference.

\noindent
The two very familar examples are:

(i) $B^0_d\to \psi K_S$, where $\xi=-1,~\phi_f={\rm Arg}(V^*_{cb}V_{cs})=0$,
\beq
{\cal A}(t)=-\sin2\beta\sin(\Delta mt)~,
\eeq
and

(ii) $B^0_d\to\pi^+\pi^-$, where $\xi=1,~\phi_f={\rm Arg}(V^*_{ub}V_{ud})=
\gamma$,
\beq
{\cal A}(t)=-\sin2\alpha\sin(\Delta mt)~.
\eeq
Thus, the two asymmetries measure the angles $\beta$ and $\alpha$.

\subsection{\it Decays to other States}

A similar method can also be applied to measure weak phases when $f$ is a
common decay  mode of $B^0$ and $\Bbar$, but not
necessarily a CP eigenstate. In this case one measures four different
time-dependent decay rates, $\Gamma_f(t),~\Gammabar_f(t),~\Gamma_{\fbar}(t),~
\Gammabar_{\fbar}(t)$, corresponding to initial $B^0$ and $\Bbar$ decaying to
$f$ and its charge-conjugate $\fbar$ \cite{MG}. The four rates depend on four
unknown quantities, $|A|,~|\Abar|,~\sin(\Delta\delta_f+\Delta\phi_f+2\phi_M),~
\sin(\Delta\delta_f-\Delta\phi_f-2\phi_M)$. ($A$ and $\Abar$ are the decay
amplitudes of $B^0$ and $\Bbar$ to $f$, $\Delta\delta_f$
and $\Delta\phi_f$ are the the strong and weak phase-differences between these
amplitudes). Thus, the four rate measurements allow a determination of the weak
CKM phase $\Delta\phi_f+2\phi_M$. This method can be applied to measure
$\alpha$ in $B^0_d\to\rho^+\pi^-$, and to measure $\gamma$ in
$B^0_s\to D^+_s K^-$ \cite{ADKD}. Other ways of measuring $\gamma$ in $B^0_s$
decays were discussed in Ref.~\cite{FLEIDU}.

\subsection{\it ``Penguin pollution"}

All this assumes that a single weak phase dominates the decay
$B^0(\Bbar)\to f$. As a matter of fact, in a variety of decay processes, such
as in $\bd\to\pi^+\pi^-$, there exists a second amplitude due to a ``penguin"
diagram in addition to the usual ``tree" diagram \cite{PEN}. As a result, CP is
also violated in the direct decay of a $B^0$, and one faces a problem of
separating the two types of asymmetries. This can only be partially
achieved through the more general time-dependence
\beq\label{GENASYM}
{\cal A}(t)=
{(1-|\Abar/A|^2)\cos(\Delta mt)-2{\rm
Im}(e^{-2i\phi_M}\Abar/A)\sin(\Delta mt)\over 1+|\Abar/A|^2}~.
\eeq
Here the $\cos(\Delta mt)$ term implies direct CP violation, and the
coefficient of $\sin(\Delta mt)$ obtains a correction from the penguin
amplitude. The two terms have a different dependence on $\Delta\delta$,
the final-state phase-difference between the tree and penguin amplitudes.
The coefficient of $\cos(\Delta mt)$ is proportional to $\sin(\Delta\delta)$,
whereas the correction to the coefficient of $\sin(\Delta mt)$ is proportional
to $\cos(\Delta\delta)$. Thus, if $\Delta\delta$ were small, this correction
might be large in spite of the fact that the $\cos(\Delta mt)$ term were too
small to be observed.

\subsection{\it Resolving Penguin Pollution by Isospin}

The above ``penguin pollution" may lead to dangerously large effects in
$B^0_d(t)\to\pi^+\pi^-$ decay, which would avoid a clean determination of
$\alpha$ \cite{PENPI}. One way of removing this effect is by measuring also the
(time-integrated) rates of $\bd\to\pi^0\pi^0$, $B^+\to\pi^+\pi^0$
and their charge-conjugates \cite{GRLO}. One uses the different isospin
properties
of the penguin ($\Delta I=1/2$) and tree ($\Delta I=1/2, 3/2$) operators and
the well-defined weak phase of the tree operator. This enables one to determine
the correction to $\sin2\alpha$ in the second term of Eq.(\ref{GENASYM}).
Electroweak penguin contributions could, in principle, spoil this method, since
unlike the QCD penguins they are not pure $\Delta I=1/2$ \cite{DH}. These
effects
are, however, very small and consequently lead to a tiny uncertainty in
determining $\alpha$ \cite{EWP}. The difficult part of this method
may perhaps be the decay rate measurement into two neutral pions. It is of
major
importance to settle experimentally the question of color-suppression of this
mode. Other methods of resolving the
``penguin pollution" in $\bd\to\pi^+\pi^-$, which do not rely on decays
to neutral pions, will be described in Sec. 3.

\subsection{\it Measuring $\gamma$ in $B^{\pm}\to D K^{\pm}$}

In $B^{\pm}\to D K^{\pm}$, where $D$ may be either a flavor state ($D^0,~
\Dbar$) or a CP-eigenstate ($D^0_1,~D^0_2$), one can measure separately the
magnitudes of two interfering amplitudes leading to direct CP violation.
This enables a measurement of $\gamma$, the relative weak phase between these
two amplitudes \cite{GW}. This method is based on a simple quantum mechanical
relation among the amplitudes of three different processes,
\beq
\sqrt{2}A(B^+\to D^0_1 K^+)~=~A(B^+\to D^0 K^+)~+~A(B^+\to \Dbar K^+)~.
\label{tr}
\eeq
The CKM factors of the two terms on the right-hand-side,
$V^*_{ub}V^{~}_{cs}$ and $V^*_{cb}V^{~}_{us}$, involve the weak phases $\gamma$
and zero, respectively. A similar triangle relation can be written for the
charge-conjugate processes. Measurement of the rates of these six proccesses,
two pairs of which are equal, enables a determination of $\gamma$.
The present upper limit on the branching ratio of $B^+\to \overline{D}^0 K^+$
\cite{DK} is already very close to the value expected in the SM. The major
difficulty of this method would be measuring $B^+\to D^0 K^+$, if this process
were color-suppressed similarly to $B^0\to\Dbar\pi^0$ \cite{PDG}. For further
details and a feasibility study see Ref.~\cite{STONE}.

\noindent
If indeed $B(B^+\to D^0 K^+)$ is found to be suppressed to a level of $10^{-6}$,
then one of the sides of the triangle Eq.(11) would be much smaller than the
other two, creating a serious difficulty in observing an asymmetry.
The other consequence of such suppression would be a difficulty in determining
the flavor of $D^0$ through its hadronic decays, which interfere with
Cabibbo-suppressed decays of $\Dbar$ from $B^+\to\Dbar K^+$. (This problem will
be addressed below). The easy way out is to compare other two processes
of this kind, again induced by $V^*_{ub}V^{~}_{cs}$ and
$V^*_{cb}V^{~}_{us}$, which are equally suppressed. Although
this does not improve statistics, the resulting CP asymmetries are expected to
be larger. Two variants, based on this simple idea, use the following processes:
\begin{itemize}
\item $B^0\to D^0(\Dbar)K^{*0}$, where the flavor of $K^{*0}$ is determined
through $K^{*0}\to K^+\pi^-$. Both decays to $D^0$ and $\Dbar$ are
color-suppressed \cite{GLD}.
\item $B^+\to D^0(\Dbar) K^+$, where $D^0$ and $\Dbar$ are identified by their
Cabibbo-allowed and Cabibbo-suppressed decays to $K^-\pi^+$, respectively.
In this case the two interfering amplitudes forming a triangle with their sum,
one being color-suppressed and the other being doubly-Cabibbo-suppressed,
may be of comparable  magnitudes \cite{ADS}.

\section{Methods based on Flavor SU(3)}

\subsection{\it $\alpha, \beta$ and $\gamma$ from $B$ Decays to two light
pseudoscalars}

One may use approximate flavor SU(3) symmetry of strong interactions, including
first order SU(3) breaking, to relate
all two body processes of the type $B\to\pi\pi,~B\to\pi K$ and $B\to K\Kbar$.
Since SU(3) is expected to be broken by effects of order $20\%$, such
as in $f_K/f_{\pi}$, one must introduce SU(3) breaking terms in such an
analysis.
This approach has recently received special attention
\cite{GHLR,WOLF,DHE,BF,KP,GL}.
In the present section we will discuss two applications of this analysis to
a determination of weak phases.
Early applications of SU(3) to two-body $B$ decays can be found in
Ref.~\cite{EARLY}.

The weak Hamiltonian operators associated with the transitions
$\bbar\to\ubar uq$
and $\bbar\to \qbar$ ($q=d$ or $s$) transform as a ${\bf 3^*},~{\bf 6}$ and
${\bf 15^*}$ of SU(3). The $B$ mesons are in a triplet, and the symmetric
product of two final state pseudoscalar octets in an S-wave contains a singlet,
an octet and a 27-plet. Thus, these processes are given in terms of
five SU(3) amplitudes: $\langle  ~1~ || ~3^* || 3 \rangle,~\langle  8 ||
~3^* || 3
\rangle,~\langle  8 || ~6~  || 3 \rangle,~\langle  8 || 15^* || 3 \rangle,~
\langle 27 || 15^* || 3 \rangle$.

An equivalent and considerably more convenient representation of these
amplitudes is given in terms of an overcomplete set of six quark diagrams
occuring in five different combinations.
These diagrams are denoted by $T$ (tree), $C$ (color-suppressed),
$P$ (QCD-penguin), $E$ (exchange), $A$ (annihilation) and $PA$ (penguin
annihilation). The last three amplitudes, in which the spectator quark enters
into the decay Hamiltonian, are expected to be suppressed by $f_B/m_B$
($f_B\approx 180~{\rm MeV}$) and may be neglected to a good approximation.

The presence of higher-order electroweak penguin contributions introduces no new
SU(3) amplitudes, and in terms of quark graphs merely leads to a substitution
\cite{EWP}
\beq \label{eqn:combs}
T\to t\equiv T + P^C_{EW}~~,~~
C\to c\equiv C + P_{EW}~~,~~
P\to p\equiv P-{1\over 3}P^C_{EW}~~,
\eeq
where $P_{EW}$ and $P^C_{EW}$ are color-favored and color-suppressed
electroweak penguin amplitudes. $\Delta S=0$ amplitudes are denoted by unprimed
quantities and $|\Delta S|=1$ processes by primed quantities. Corresponding
ratios are given by ratios of CKM factors
\beq
{T'\over T} = {C'\over C} = {V_{us}\over V_{ud}}~,~~~~~~{P'\over P} =
{P'_{EW}\over P_{EW}} = {V_{ts}\over V_{td}}~.
\eeq
$t$-dominance was assumed in the ratio $P'/P$. The effect of $u$ and $c$ quarks
in penguin amplitudes can sometimes be important \cite{BUFLE}.

The expressions of all thirteen two body decays to two light pseudoscalars in
the SU(3) limit are given in Tables 1 and 2 of \cite{EWP}.
The vanishing of three other amplitudes, associated with $B^0_d\to K^+ K^-,~
B^0_s\to\pi^+\pi^-,~B^0_s\to\pi^0\pi^0$, follows from the assumption of
negligible exchange ($E$) amplitudes. This can be used to test our assumption
which neglects final state rescattering effects. If rescattering is important,
then the rates of the above processes could be considerably larger than
estimated using na\"{\i}ve factorization. This possibilty will be discussed in
Section 4.

First-order SU(3) breaking corrections can be introduced in a most general
manner through parameters describing mass insertions in the above quark
diagrams \cite{SU3BR}. The interpretation of these corrections in terms of
ratios of decay constants and form factors is model-dependent. There is,
however, one  case in which such interpretation is quite reliable. Consider
the tree amplitudes $T$ and $T'$. In $T$ the $W$ turns into a $u\dbar$ pair,
whereas in $T'$ it turns into $u\sbar$. One may assume factorization for
$T$ and
$T'$, which is supported by data on $B\to \overline{D}\pi$ \cite{BS}, and is
justified for $B\to\pi\pi$ and $B\to\pi K$ by the high momentum with which
the two color-singlet mesons separate from one another. Thus,
\beq\label{SU3BRK}
{T'\over T} = {V_{us}\over V_{ud}}{f_K\over f_{\pi}}~.
\eeq
Similar assumptions for $C'/C$ and $P'/P$ cannot  be justified.

Tables 1 and 2 of \cite{EWP} and Eq.(\ref{SU3BRK}) can be used to separate the
penguin term from
the tree amplitude in $B^0_d\to\pi^+\pi^-$, and thereby determine
simultaneously
all the three angles of the unitarity triangle. In one of the schemes
\cite{MGJR} one uses only $B$ decays to final states with kaons and charged
pions, $B^0_d\to\pi^+\pi^-,~B^0_d\to\pi^- K^+,~B^+\to\pi^+ K^0$ and the
corresponding charge-conjugated processes. Measurement of these six processes
enables a determination of both $\alpha$ and $\gamma$, with some remaining
discrete ambiguity associated with the size of final-state phases. A sample
corresponding to about 100 $B^0_d\to\pi^+\pi^-$ events, 100 $B^0_d\to
\pi^{\pm}K^{\mp}$ events, and a somewhat smaller number of detected
$B^{\pm}\to\pi^{\pm}K_S$ events, attainable in future $e^+e^-$
$B$-factories, is
sufficient for reducing the presently allowed region in the ($\alpha,~\gamma$)
plane by a considerable amount. The reader is referred to Ref.30
for more details. A few alternative ways to learn the penguin effects in
$B^0_d\to\pi^+\pi^-$ were suggested in Ref.31.

\subsection {\it Use of the Recently Observed $B\to\eta' K$}

The use of $\eta$ and $\eta'$ allows a determination of $\gamma$ from decays
involving charged $B$ decays alone \cite{ETA}.
When considering final states involving $\eta$ and $\eta'$ one encounters an
additional penguin diagram (a so-called ``vacuum cleaner" diagram),
contributing
to decays involving one or two flavor SU(3) singlet pseudoscalar mesons
\cite{DGR2}. This amplitude ($P_1$) appears in a fixed combination with a
higher-order electroweak penguin contribution in the form $p_1\equiv P_1-
(1/3)P_{EW}$. The importance of this diagram was demonstrated very recently by
the large branching ratio $B(B^+\to\eta' K^+)=(7.8^{+2.7}_{-2.2}\pm 1.0)\times
10^{-5}$ reported at this conference \cite{KIM}.

Writing the physical states in terms of the SU(3) singlet and
octet states
\beq
\eta=\eta_8\cos\theta - \eta_1\sin\theta,~~~\eta'=\eta_8\sin\theta+\eta_1
\cos\theta,~~~\sin\theta\approx {1\over 3},
\eeq
one finds the following expressions for the four possible $\Delta S=1$
amplitudes of charged $B$ decays to two charmless pseudoscalars:
$$
A(B^+\to \pi^+ K^0)=p'~,~~~~A(B^+\to\pi^0 K^+)={1\over \s}(-p'-t'-c')~,
$$
$$
A(B^+\to\eta K^+)={1\over\sqrt{3}}(-t'-c'-p'_1)~,
$$
\beq
A(B^+\to \eta' K^+)=
{1\over\sqrt{6}}(3p'+t'+c'+4p'_1)~.
\eeq
These amplitudes satisfy a quadrangle relation
$$
\sqrt{6}A(B^+\to \pi^+ K^0) + \sqrt{3}A(B^+\to\pi^0 K^+)
$$
\beq
- 2\sqrt{2}A(B^+\to\eta K^+) - A(B^+\to \eta' K^+) = 0~.
\eeq

A similar quadrangle relation is obeyed by the charge-conjugate amplitudes, and
the relative orientation of the two quadrangles holds information about
weak phases. However, it is clear that each of the two quadrangles cannot be
determined from its four sides given by the measured amplitudes. A closer look
at the expressions of the amplitudes shows that the two quadrangles share a
common base, $A(B^+\to \pi^+ K^0)=A(B^-\to \pi^- \overline{K}^0)$, and the two
sides opposite to the base (involving $\eta$) intersect at a point lying
3/4 of the distance from one vertex to the other. This fixes the shapes of
the quadrangles up to discrete ambiguities. Finally, the phase $\gamma$ can be
determined by relating these amplitudes to that of $B^+\to\pi^+\pi^0$
$$
\vert A(B^+\to\pi^0 K^+) -  A(B^-\to\pi^0 K^-)\vert
$$
\beq
= 2{V_{us}\over V_{ud}}{f_K\over f_{\pi}}\vert
A(B^+\to\pi^+\pi^0)\vert\sin\gamma~.
\eeq
This method becomes particularly appealing due to the recent CLEO
measurement of an anomalously large branching ratio of
$B^+\to \eta' K^+$ \cite{KIM}.
\section{Large Final State phases in $B$ Decays}

In order to have large asymmetries in charged $B$ decays one requires an
interference between two amplitudes of comparable magnitude, involving both a
large weak CKM phase-difference and a large final state interaction
phase-difference. So far, there exists no experimental evidence for final state
phases in $B$ decays, and it has been often assumed that such phases are likely
to be small in decays of a heavy $B$ meson to two light high momentum particles.
Evidence for strong phases, related to final states with well-defined isospin
and angular momentum, can be obtained from $B\to\overline{D}\pi$ decays. The
amplitudes into $D^-\pi^+,~\overline{D}^0\pi^0,~\overline{D}^0\pi^+$ obey a
triangle relation, from which the phase-difference between the $I=1/2$ and
$I=3/2$ amplitudes may be determined. The present branching ratios of these
decays already imply an upper limit \cite{YAMA}, $\delta_{1/2}-
\delta_{3/2}<35^{\circ}$. Improved measurements of these braching ratios may
lead to first evidence for strong phases or to more stringent bounds.
An important question is, therefore, where would one expect
final state interaction phases to be large?
In the present section we will demonstrate two cases in which large phases may
be anticipated.
\subsection{\it Interference between Resonance and Background}

Consider the decay $B^+\to\chi_{c0}\pi^+,~\chi_{c0}\to\pi^+\pi^-$, where one is
looking for a final state with three pions, two of which have an invariant
mass around $m(\chi_{c0})=3415$ MeV \cite{EGM}. The width of this
$J^P=0^+$~$c\cbar$ state,
$\Gamma(\chi_{c0})=14\pm 5$ MeV, is sufficiently large to provide a large, and
probably maximal, CP conserving phase.
The decay amplitude into three pions, where two pions are at the resonance,
consists of two terms with different CKM phases (we neglect a small penguin
term):
\item{R=} a resonating amplitude, consisting of a product of the weak decay
amplitude of $B^+\to\chi_{c0}\pi^+$ involving a real CKM factor
$V^*_{cb}V_{cd}$  ($a_w$=real), the strong decay amplitude of
$\chi_{c0}\to\pi^+\pi^-$ ($a_s$=real), and a
Breit-Wigner term for the intermediate $\chi_{c0}$.
\item{D=} a direct decay amplitude of $B^+\to\pi^+\pi^-\pi^+$ involving a CKM
factor $V^*_{ub}V_{ud}$ with phase $\gamma$, which we write as
$(d/m_B)\exp(i\gamma)$ (d=real):
\beq
R=a_wa_s{\sqrt{m\Gamma}\over s-m^2+im\Gamma}~,~~~~~~D={d\over
m_B}\exp(i\gamma)~.
\eeq
The total amplitude is $R+D$.

The $B^+~-~B^-$ decay rate asymmetry, integrated symmetrically around the
resonance,
is given by
\beq
Asym.\approx -2{d\over a_wa_s}{\sqrt{m\Gamma}\over m_B}\sin\gamma~.
\eeq
The strong phase difference between the resonating and direct amplitudes is
approximately $\pi/2$. Phases other than due to the resonance width were
neglected.
Reasonable estimates of the amplitudes $d$ and $a_wa_s$ show that the
coefficient of $\sin\gamma$ in the asymmetry is of order one \cite{EGM}.
That is, a large CP asymmetry is expected in this channel, requiring for its
observation $10^8$ to $10^9$ $B$ mesons.
\subsection{\it Rescattering in Quark Annihilation Processes}

A large number of $B$ meson decays may proceed only through participation of
the spectator quark, whether through amplitudes proportional to $f_B/m_B$ or
via rescattering from other less-suppressed amplitudes. A recent analysis of
this class of processes was carried out \cite{BGR}, assuming that rescattering
from a dominant process leads to suppression by only factor
$\lambda\sim 0.2$ compared to $f_B/m_B\approx\lambda^2$. Such an assumption
can be justified, for instance, by a Regge-based analysis \cite{BH}.
The consequences of this assumption are twofold:
\item An expected hierarchy of amplitudes in the absence of rescattering will
be violated by rescattering corrections, leading to much larger rates. As an
example, the branching ratio of $B^0 \to K^+ D_s^-$ can be enhanced by
rescattering through a $\pi^+ D^-$ intermediate state from about $10^{-6}$ to
somewhat less than $10^{-4}$.
\item Such violations could point the way toward channels in which
final-state interactions could be important. Cases in which final state phases
lead to large CP asymmetries are those to which both tree and penguin
amplitudes
contribute. Two examples are $B^0 \to D^0 \bar D^0 (D^+_s D^-_s)$ and
$B_s \to\pi^+\pi^- (\pi^0\pi^0)$.
\section{CP Violation Beyond the Standard Model}

\subsection{\it Modifying the Unitarity Triangle}

The above discussion assumes that the only source of CP violation is the phase
of the CKM matrix. Models beyond the SM involve other phases, and consequently
the measurements of CP asymmetries may violate SM constraints on the three
angles of the unitarity triangle \cite{DLN}. Furthermore, even in the absence
of new CP
violating phases, these angles may be affected by new contributions to the
sides of the triangles. The three sides, $V_{cd}V^*_{cb},
~V_{ud}V^*_{ub}$ and $V_{td}V^*_{tb}$, are measured in $b\to c l\nu,~
b\to u l\nu$ and in $\bd-\bdb$ mixing, respectively.
A variety of models beyond the SM provide new contributions to $\bd-\bdb$ and
$\bs-\bsb$ mixing, but only very rarely \cite{WAKA} do such models involve new
amplitudes which can compete with the $W$-mediated tree-level $b$ decays.
Therefore, whereas two of the sides of the unitarity triangle are usually
stable under new physics effects, the side involving $V_{td}V^*_{tb}$ can be
modified by such effects. In certain models, such as a four generation model
and models involving $Z$-mediated flavor-changing neutral currents (to be
discussed below), the unitarity triangle turns into a quadrangle.

In the phase convention of Eq.(\ref{V})
the three angles $\alpha,~\beta,~\gamma$ are defined as follows:
\beq
\gamma\equiv {\rm Arg}(V_{ud}V^*_{ub})~,~~~~~~ \beta\equiv {\rm Arg}
(V_{tb}V^*_{td})~,~~~~~~\alpha\equiv \pi-\beta-\gamma~.
\eeq
Assuming that new physics affects only $B^0-\overline{B}^0$
and $\bs-\bsb$ mixing, one can make the
following simple observations about CP asymmetries beyond the SM:
\item The asymmetry in $B^0_d\to\psi K_S$ measures the phase of $\bd-\bdb$
mixing and is defined as $2\beta'$, which in general can be different from
$2\beta$.
\item  The asymmetry in $B^0_d\to\pi^+\pi^-$ measures the phase of $\bd-\bdb$
mixing plus twice the phase of $V^*_{ub}$, and is given by $2\beta'+
2\gamma\equiv 2\pi-2\alpha'$, where $\alpha'\ne\alpha$.
\item The time-dependent rates of $\bs/\bsb\to D^{\pm}_s K^{\mp}$ determine a
phase $\gamma'$ given by the phase of $\bs-\bsb$ mixing plus the phase of
$V^*_{ub}$; in this case $\gamma'\ne\gamma$.
\item The processes $B^{\pm}\to D^0 K^{\pm},~B^{\pm}\to \Dbar K^{\pm},
~B^{\pm}\to D^0_{1(2)} K^{\pm}$ measure the phase of $V^*_{ub}$ given by
$\gamma$.

Measuring a nonzero value for the phase $\gamma$ through the last method
would be evidence for CP violation in direct decay, thus ruling out
superweak-type models \cite{SUPER}. Such a measurement will
obey the triangle relation $\alpha'+\beta'+\gamma=\pi$
with the phases of $B^0_d\to\psi K_S$ and $B^0_d\to\pi^+\pi^-$, irrespective of
contributions from new physics to $B^0-\Bbar$ mixing. On the other
hand, the phase $\gamma'$ measured by the third method violates this relation.
This demonstrates the importance of measuring phases in a variety of
independent
ways.

Another way of detecting new physics effects is by determining the phase
of the $\bs-\bsb$ mixing amplitude which is extremely small in the SM,
corresponding to an angle of the almost flat $sb$ unitarity triangle
\cite{LOWOL}. This can be achieved through CP asymmetry measurements in
decays such as $\bs\to\psi\phi$ governed by the quark process $b\to c\cbar s$.

Let us note in passing that in certain models, such as multi-Higgs doublet
models with natural flavor conservation (to be discussed below), in spite
of new
contributions to $\bd-\bdb$ and $\bs-\bsb$ mixing, the phases measured in
$B^0_d\to\psi K_S$ and in $\bs/\bsb\to D^{\pm}_s K^{\mp}$ are unaffected,
$\beta'=\beta,~\gamma'=\gamma$. Nevertheless, the values measured for these
phases may be inconsistent with the CP conserving measurements of the sides of
the unitarity triangle.

\subsection {\it CP Asymmetries vs. Penguin Decays}

Models in which CP asymmetries in $B$ decays are affected by new contributions
to $B^0-\overline{B}^0$ mixing will usually also have new amplitudes
contributing
to rare flavor-changing $B$ decays, such as $b\to s X$ and $b\to d X$.
We refer to such processes, involving a photon, a pair of leptons or hadrons
in the final state, as ``penguin" decays.

In the SM both $B^0-\overline{B}^0$ mixing and penguin decays are governed by
the CKM parameters $V_{ts}$ and $V_{td}$. Unitarity of the CKM matrix implies
\cite{ALON} $\vert V_{ts}/V_{cb}\vert\approx 1$, $0.11 < \vert V_{td}/V_{cb}
\vert < 0.33$, and $\bd-\bdb$ mixing only improves the second constraint
slightly due to large hadronic uncertainties,
$0.15 < \vert V_{td}/V_{cb}\vert < 0.33$.

The addition of contributions from new physics to $B^0-\overline{B}^0$ mixing
relaxes the above constraints in a model-dependent manner. The new
contributions
depend on new couplings and new mass scales which appear in the models. These
parameters also determine the rate of penguin decays. A recent comprehensive
model-by-model study \cite{MGDL}, updating previous work, showed that the
values of the new
physics parameters, which yield significant effects in $B^0-\overline{B}^0$
mixing, will also lead in a variety of models to large deviations from the SM
predictions for certain penguin decays. Here we wish to briefly summarize
the results of this analysis:

\item {\it Four generations}: The magnitude and phase of $B^0-\overline{B}^0$
mixing
can be substantially changed due to new box-diagram contributions involving
internal $t'$ quarks. For such a region in parameter space, one expects an
order-of-magnitude enhancement (compared to the SM prediction) in the branching
ratio of $\bd\to l^+l^-$ and $B^+\to\phi\pi^+$.
\item {\it $Z$-mediated flavor-changing neutral currents}: The magnitude and
phase of $B^0-\overline{B}^0$ mixing can be altered by a tree-level
$Z$-exchange.
If this effect is large, then the branching ratios of the penguin processes
$b\to s l^+ l^-,~\bs\to l^+\l^-,~\bs\to\phi\pi^0~(b\to d l^+ l^-,~
\bd\to l^+\l^-,~B^+\to\phi\pi^+$) can be enhanced by as much as one (two)
orders-of-magnitude.
\item {\it Multi-Higgs doublet models with natural flavor conservation}:
New box-diagram contributions to $B^0-\overline{B}^0$ mixing with internal
charged
Higgs bosons affect the magnitude of the mixing amplitude but not its phase
(measured, for instance, in $\bd\to\psi K_S$). When this effect is large, the
branching ratios of $\bd,\bs\to l^+ l^-$ are expected to be larger than in
the SM
by up to an order of magnitude.
\item {\it Multi-Higgs doublet models with flavor-changing neutral scalars}:
Both the magnitude and phase of $B^0-\overline{B}^0$ mixing can be changed
due to a
tree-level exchange of a neutral scalar. In this case one expects no
significant
effects in penguin decays.
\item {\it Left-right symmetric models}: Unless one fine-tunes the right-handed
quark mixing matrix, there are no significant new contributions in
$B^0-\overline{B}^0$ mixing and in penguin $B$ decays.
\item {\it Minimal supersymmetric models}: There are a few new contributions to
$B^0-\overline{B}^0$ mixing, all involving the same phase as in the SM.
Branching
ratios of penguin decays are not changed significantly. However, certain energy
asymmetries, such as the $l^+l^-$ energy asymmetry in $b\to s l^+ l^-$ can be
largely affected.
\item{\it Non-minimal supersymmetric models}: In non-minimal SUSY models with
quark-squark alignment, the SUSY contributions to $B^0-\overline{B}^0$ mixing
and to penguin decays are generally small. In other models, in which all SUSY
parameters are kept free, large contributions with new phases can appear in
$B^0-\overline{B}^0$ mixing and can affect considerably SM predictions for
penguin
decays. However, due to the many parameters involved, such schemes have little
predictivity.

We see that measurements of CP asymmetries and rare penguin decays
give complementary information and, when combined, can distinguish among the
different models. In models of the first, second and fourth types one has
$\beta'\ne\beta,~\gamma'\ne\gamma$. One expects different measurements of
$\gamma$ in
$B^{\pm}\to D K^{\pm}$ and in $\bs/\bsb\to D_s^{\pm} K^{\mp}$, and a nonzero
CP asymmetry in $\bs\to\psi\phi$. These three models can then be distinguished
by their different predictions for branching ratios of penguin decays.
On the other hand, both in the third and sixth models one expects
$\beta'=\beta,~\gamma'=\gamma$. In order to distinguish between these two
models, one would
have to rely on detailed dilepton energy distributions in $b\to s l^+ l^-$.

\subsection{\it Large CP Asymmetries in Radiative Neutral $B$ Decays}

Certain CP asymmetries, such as in $B_s \to J/\psi\phi$, are expected to be
extremely small in the SM, and are therefore very sensitive to sources of CP
violation beyond the SM. This is a typical case, in which large effects of new
physics in CP asymmetries originate in additional
sizable contributions to $\bq-\bqb~(q=d,s)$ mixing \cite{MGDL}. Much smaller
effects, which are harder to measure and have considerable theoretical
uncertainties, can occur as new contributions to $B$ decay amplitudes
\cite{GWOR}. There is one class of processes, namely radiative
$\bo$ and $\bs$ decays, in which large mixing-induced
asymmetries are due to new contributions to the decay amplitude \cite{AGS}.

Consider decays of the type $\bo,\bs \to M^0\gamma$, where $M^0$ is any hadronic
self-conjugate state $M^0 = \rho^0,\omega,\phi,K^{*0}$~(where $K^{*0}\to K_S
\pi^0$), etc. As in $B^0 \to J/\psi K_S$ and $B_s \to J/\psi\phi$, the
asymmetries in $B\to M^0\gamma$ are due to the interference between
mixing and decay. We neglect direct CP violation which is expected to be small
\cite{DIR}.
In the Standard Model, the photon in $b\to q\gamma$ is dominantly left-handed;
only a fraction $m_q/m_b$ of the amplitude corresponds to a right-handed photon,
where the quark masses are current masses. The final $M^0\gamma$ states are not
pure CP-eigenstates; they consist to a good approximation (neglecting the ratio
$m_q/m_b$) of equal admixtures of states with positive and negative
CP-eigenvalues. Thus, due to an almost complete cancellation between
contributions from positive and negative CP-eigenstates, the asymmetries in
$b\to q\gamma$ are very small, given by $m_q/m_b$.  A few examples of
time-dependent asymmetries expected in the SM are \cite{AGS}:
\begin{eqnarray}
\bo \to K^{*0}\gamma &:& {\cal A}(t) \approx (2m_s/m_b)\sin(2\beta)
\sin(\Delta mt)~,
\nonumber\\
\bo \to \rho^0\gamma &:& {\cal A}(t) \approx 0~,
\nonumber\\
\bs \to \phi\gamma &:& {\cal A}(t) \approx 0~,
\nonumber\\
\bs \to K^{*0}\gamma &:& {\cal A}(t) \approx -(2m_d/m_b)\sin(2\beta)
\sin(\Delta mt)~,
\label{examples}
\end{eqnarray}
where $K^{*0}$ is observed through $K^{*0}\to K_S \pi^0$.

Much larger CP asymmetries can occur in extensions of the Standard Model
such as the
$SU(2)_L\times SU(2)_R\times U(1)$ left-right symmetric model \cite{MOHAP},
$SU(2)\times U(1)$ models with exotic fermions (mirror or vector-doublet quarks)
\cite{LANLON}, and nonminimal supersymmetric models \cite{MASIERO}. As an
example, consider the left-right symmetric model, in which mixing of
the weak-eigenstates $W_L, W_R$ into the mass-eigenstates $W_1, W_2$ is given
by
\beq
\left( \begin{array}{c}
W_1^+\\
W_2^+
\end{array} \right)
=
\left( \begin{array}{cc}
\cos\zeta & e^{-i\omega}\sin\zeta \\
-\sin\zeta & e^{-i\omega}\cos\zeta
\end{array} \right)
\ \ \
\left( \begin{array}{c}
W_L^+\\
W_R^+
\end{array} \right)~.
\eeq
The process $b \to q\gamma$ obtains in addition to the SM penguin amplitude
with $W$ (and $t$) exchange, two penguin-type contributions, from $W_L-W_R$
mixing and from charged scalar exchange. These terms can be sizable in spite of
present severe constraints on the parameters of the model, the $W_R$ and
charged Higgs masses and the mixing angle $\zeta$. The measured branching ratio
of $b\to s\gamma$ \cite{CLEO} implies certain constraints on these parameters.
However, even if experiments were to agree precisely with the SM prediction,
the asymmetrrie could be very large. For this case, we list the largest possible
asymmetries in the processes of Eqs.(22), obtained when $\zeta$
takes its present experimental upper limit $\zeta=0.003$ \cite{AGS},
\begin{eqnarray}
\bo \to K^{*0}\gamma &:& {\cal A}(t) \approx \mp 0.67\cos(2\beta)
\sin(\Delta mt)~,
\nonumber\\
\bo \to \rho^0\gamma &:& {\cal A}(t) \approx \mp 0.67\cos(2\beta)
\sin(\Delta mt)~,
\nonumber\\
\bs \to \phi\gamma &:& {\cal A}(t) \approx \mp 0.67\sin(\Delta mt)~,
\nonumber\\
\bs \to K^{*0}\gamma &:& {\cal A}(t) \approx \mp 0.67\sin(\Delta mt)~.
\label{examplesLR}
\end{eqnarray}
That is, whereas in the SM all asymmetries are at most a few percent, they
can be larger than 50$\%$ in the left-right symmetric model. Observing
asymmetries at this level would be a clear signal of physics
beyond the SM.

\section{Conclusion: Future Outlook}

The three main goals of future $B$ physics experiments, related to CP violation
as discussed above, are the following:
\item Observation of CP asymmetries in $B^0,~B^+,~B_s$ decays would
provide first evidence for CP violation outside the neutral kaon system.
\item Determination of $\alpha,~\beta,~\gamma$ from these asymmetries and
from $B$ decay rates would be complementary to information about the  unitarity
triangle from CP conserving measurements and from CP violation in the neutral
$K$ meson system.
\item Detection of deviations from Standard Model asymmetry predictions,
combined with information about rates of rare penguin $B$ decays,
could provide clues to a more complete theory.
\bigskip

\noindent
{\bf Acknowledgements}

\noindent
This work was supported in part by the United
States-Israel Binational Science Foundation under Research Grant Agreement
94-00253/2, and by the Israel Science Foundation.
\newpage

\def \np#1#2#3{Nucl.~Phys.~B{\bf#1}, #2 (#3)}
\def \plb#1#2#3{Phys.~Lett.~B{\bf#1}, #2 (#3)}
\def \prd#1#2#3{Phys.~Rev.~D {\bf#1}, #2 (#3)}
\def \prl#1#2#3{Phys.~Rev.~Lett.~{\bf#1}, #2 (#3)}
\def \prp#1#2#3{Phys.~Rep.~{\bf#1} #2 (#3)}
\def \ptp#1#2#3{Prog.~Theor.~Phys.~{\bf#1}, #2 (#3)}
\def \rmp#1#2#3{Rev.~Mod.~Phys.~{\bf#1} #2 (#3)}
\def \zpc#1#2#3{Z.~Phys.~C {\bf#1}, #2  (#3)}
\def \ite{{\it et al.}}
\def \stone{{\it B Decays}, edited by S. Stone (World Scientific, Singapore,
1994)}

\end{itemize}
\end{document}